\documentstyle[epsfig,aps,prb,amstex,graphicx]{revtex}
\newcommand\beq{\begin{equation}}
\newcommand\eeq{\end{equation}}
\newcommand\bea{\begin{eqnarray}}
\newcommand\eea{\end{eqnarray}}

\begin{document}
\baselineskip=16pt
\begin{center}
{\large {\bf Model Exact Low-Lying States and Spin Dynamics in Ferric Wheels; 
Fe$_6$ to Fe$_{12}$}}\\
{\bf Indranil Rudra$^1$, S. Ramasesha$^1$ and Diptiman Sen $^2$\\}
$^1$Solid State and Structural Chemistry Unit\\
Indian Institute of Science, Bangalore 560 012, India\\
$^2$ Centre for Theoretical Studies\\
Indian Institute of Science, Bangalore 560 012, India\\
\end{center}

\begin{abstract}
Using an efficient numerical scheme that exploits spatial symmetries and 
spin-parity, we have obtained the exact low-lying eigenstates of exchange
Hamiltonians for ferric wheels up to Fe$_{12}$. The largest calculation involves
the Fe$_{12}$ ring which spans a Hilbert space dimension of about 145 million
for M$_s$=0 subspace. Our calculated gaps from the singlet ground state to 
the excited triplet state agrees well with the experimentally measured values.
Study of the static structure factor shows that the ground state is 
spontaneously dimerized for ferric wheels. Spin states of ferric wheels can be
viewed as quantized states of a rigid rotor with the gap between the ground and
the first excited state defining the inverse of moment of inertia. 
We have studied the
quantum dynamics of Fe$_{10}$ as a representative of ferric wheels. We use the 
low-lying states of Fe$_{10}$ to solve exactly the time-dependent Schr\"odinger
equation and find the magnetization of the molecule in the presence of an 
alternating magnetic field at zero temperature. We observe a nontrivial 
oscillation of magnetization which is dependent on the amplitude of the {\it ac}
field. We have also studied the torque response of Fe$_{12}$ as a function of
magnetic field, which clearly shows spin-state crossover. 

\end{abstract}

\vskip .5 true cm

\hspace{3cm} PACS numbers: ~75.45.+j,~75.50.Xx,~75.60.Ej 

\section{Introduction}
In recent years polyoxometalates which are practical realization of 
nanomagnets have become the area of intense research because of their 
enormous variety of structure and fascinating magnetic properties. 
A particular aesthetic class is that of the ring-shaped iron(III) 
compounds denoted as molecular ferric wheels. The decanuclear wheel 
Fe$_{10}$ synthesized by Lippard et. al. \cite{lippard} may be regarded
as a prototype of this class. Meanwhile synthesis of ferric wheels with 
6, 8, 12 and 18 have also been reported \cite{synth}. These materials 
have dominant anti-ferromagnetic coupling between ${\rm Fe(III)}$ spins 
and a singlet ground state. The magnetic properties of such nanoscopic 
molecules result from the interplay of superexchange interactions between 
the atomic spins, dipolar coupling of the local moments and on-site 
anisotropies arising from ligand configurations. The emergence of ferric 
wheels has led to a renewed interest in the properties of the Heisenberg 
chain, especially for large spin values. In 1964 Fisher used a classical 
treatment \cite{fisher}, in which spin quantization is absent, to study 
the properties of Heisenberg Hamiltonian and found analytical solutions 
for the thermodynamic properties of the system. Numerous quantum 
mechanical calculations have been made for Heisenberg Hamiltonian, 
mostly for spin-1/2 chains. Calculations for larger ${\rm S}$ became 
more interesting after Haldane's conjecture \cite{haldane} regarding 
the presence of an energy gap in the excitation spectrum from the ground 
state for integer ${\rm S}$ chains.

\noindent
To fit the experimental temperature dependence of the magnetic susceptibility
data for ${\rm Fe_{10}}$ Lippard et. al. adopted a classical spin treatment 
to obtain the value of exchange interaction parameter ${\rm J}$. But below 
50 K this treatment fails. Ferric wheels with S=5/2, and system size of up to 
8 sites have been treated exactly using irreducible tensor operator technique 
\cite{gatteschi} with the aid of point group symmetry as well as by using
the invariance of spin Hamiltonian with regard to interchange of spin sites 
\cite{waldmann}. Recently there have been theoretical studies to explain the
low temperature magnetic susceptibility data of ${\rm Fe_{10}}$ \cite{silbey}.
The magnetization of ferric wheels exhibit a step like field dependence at low
temperature due to the occurrence of field induced ground state level crossing,
a spectacular manifestation of quantum size effect in these nanomagnets 
\cite{lippard,cornia}. There have been NMR and specific heat studies of these
ferric wheels to investigate the energy level structure \cite{julien,affronte}. 
In appropriate parameter regime these ring systems are considered to be the 
candidates for the observation of macroscopic quantum phenomena, in the form 
of quantum coherent tunneling of the Neel vector \cite{loss1}. To understand  
these low temperature properties of ferric wheels in detail we need to know 
the low lying eigenvalue spectrum for these systems. It is also of interest 
to compare and contrast the zero temperature density of state of S=5/2 rings 
with that of the exactly known S=1/2 chain. In this paper we have used spin 
parity together with rotational symmetry of the ferric wheels to obtain the 
low-lying eigenvalue spectrum for rings with 6, 8, 10 and 12 spin-5/2. The
dispersion spectrum reveals interesting features. There have been recent 
reports of solving the low-lying eigen-spectrum of ${\rm Fe_{10}}$ using the 
Density Matrix Renormalization Group \cite{loss2}. To the best of our knowledge 
ours is the first study of of ferric wheels using exact methods up to a ring 
size of 12, S=5/2 iron(III) ions.

We have also studied the quantum spin dynamics of ferric wheels in presence of 
an alternating field after setting up the Hamiltonian matrix in the subspace
of low-lying states. This Hamiltonian includes multipolar spin-spin interaction
terms besides a time varying magnetic field (Zeeman term). We have then evolved
an initial state, which is taken to be the ground state with a specific value 
of $M_S$ (the $z$-component of the total spin) in the absence of the magnetic 
field, by using the time-dependent formulation of the problem in the restricted
subspace. We observe a nontrivial oscillation of the magnetization whose 
frequency depends on the amplitude of the alternating field. This phenomena
is very similar to what has been observed in case of uniaxial magnets 
\cite{raedt}.

\section{Model Hamiltonian and Computational Details}

\subsection{Symmetry Adaptation of Correlated States}

The full symmetry of electronic states is of central interest in quantum 
theory. Since nonrelativistic Hamiltonians $\rm \hat H$ does not depend 
explicitly on spin, molecular eigenstates can be labeled by the total spin 
${\rm S}$ and the appropriate irreducible representation of the point group. 
The model Hamiltonian employed in this study is the isotropic exchange 
Hamiltonian involving exchange interaction between nearest neighbors,
\begin{equation}
{\rm 
  {\hat H} ~ = ~ \underset{<ij>}{\Sigma}J_{ij} {\hat s}_i \cdot {\hat s}_j
}
\label{ham1}
\end{equation}
where the exchange interaction ${\rm J_{ij}}$ takes the values dictated by
experimental studies of structure and magnetic properties. For ${\rm Fe_6}$ 
the $J$ value depends on the central alkali metal atom : for Na:${\rm Fe_6}$ 
$J$=32.77 K whereas for Li:${\rm Fe_6}$ $J$=20.83 K. In ${\rm Fe_{8}}$, 
${\rm Fe_{10}}$ and ${\rm Fe_{12}}$ $J$ is 22 K,15.56 K and 31.97 K 
respectively \cite{gatteschi2}. 

The total dimensionality of the Fock space of the ferric wheel is given by
\begin{equation}
{\rm 
  D_F ~=~ \overset{n}{\underset{1}{\Pi}} (2S_i+1)
}
\end{equation}
where 'n' is the total number of spins in the wheel and ${\rm S_i}$ is the 
spin on each ion. In case of ${\rm Fe_{10}}$, there are 10 iron(III) ions 
with the dimensionality of the Fock space 60,466,176. In case of ${\rm 
Fe_{12}}$, which we have solved exactly this dimension is 2,176,782,336. 

Specializing to given total $M_S$ leads to Hilbert space dimensionalities 
which are lower than the Fock space dimensionality. In the case of the ${\rm 
Fe_{12}}$ cluster the $M_S=0$ space has a dimensionality of about 145 million 
(144,840,476). The major challenge in exact computation of the eigenvalues, 
and properties of these spin clusters lies in handling such large bases and 
the associated matrices. While the dimensions look overwhelming, the matrices 
that represent the operators in these spaces are rather sparse. Usually, the 
number of nonzero elements in a row is of the order of the number of exchange 
constants in the Hamiltonian. This sparseness of the matrices allows one to 
handle fairly large systems. However, in the case of spin problems, generating 
the bases states and using the symmetries of the problem is nontrivial.

The isotropic exchange Hamiltonians conserve total spin, S, besides the
z-component of the total spin, ${\rm M_S}$. Besides these symmetries, the 
geometry of the cluster also leads to spatial symmetries which can often be 
exploited. The simplest way of generating bases functions which conserve total 
spin is the VB method that employs the Rumer-Pauling rule \cite{vb}. It is 
quite easy to generalize the Rumer-Pauling rules to a cluster consisting of 
objects with different spins to obtain states with desired total spin, S. 
However, setting-up the Hamiltonian matrix in such a basis can be 
computationally intensive since the exchange operators operating on a "legal" 
VB diagram (diagram that obeys Rumer-Pauling rules) could lead to "illegal" VB 
diagrams and resolving these "illegal" VB diagrams into "legal" diagrams would 
present the major bottle-neck. Indeed, the same difficulty is encountered when 
spatial symmetry operators operate on a VB function. Thus, the extended VB 
methods are not favored whenever one wishes to apply it to a motley 
collection of spins or when one wishes to exploit some general spatial 
symmetries that may exist in the cluster \cite{ramasesh}.

It is advantageous to partition the spaces into different total spin spaces 
because of the usually small energy gaps between total spin states which 
differ in S by unity. To avoid the difficulties involved in working with 
total spin eigenfunctions, we exploit parity symmetry in the systems. The 
parity operation involves changing the z-component of all the spins in the 
cluster from ${\rm M_{S_i}}$ to ${\rm -M_{S_i}}$. There is an associated phase 
factor with this operation given by ${\rm (-1)^{ S_{tot} + \sum_i S_i}}$. The 
isotropic exchange operator remains invariant under this operation. If this 
symmetry is employed in the ${\rm M_S=0}$ subspace, the subspace is divided 
into "even" and "odd" parity spaces depending upon the sign of the character 
under the irreducible representation of the parity group. The space which 
corresponds to even (odd) total spin we call the even (odd) parity space. 
Thus, employing parity allows partial spin symmetry adaptation which separates 
successive total spin spaces, without introducing the complications encountered 
in the VB  bases. However, the VB method can lead to complete factorization 
of the spin space leading to smaller complete subspaces.

In the ferric wheels, besides spin symmetries, there also exists spatial 
symmetries. The topology of the exchange interaction leads to a ${\rm C_n}$ 
point group symmetry, where 'n' is the number of iron ions in the ring. Hence,
${\rm Fe_{6}}, {\rm Fe_{8}}, {\rm Fe_{10}}, {\rm Fe_{12}}$ will have C$_6$,
C$_8$, C$_{10}$ and C$_{12}$ symmetry respectively. It should be mentioned that
among these ferric wheels only ${\rm Fe_{10}}$ is strictly having a ten fold
rotational symmetry, rest of them approximately have the above mentioned 
symmetry. For computational advantage  we have assumed the rotational symmetry
for all the ferric wheels. This point group appears at first site to present
difficulties because the characters in the irreducible representation are
in some cases complex which could lead to complex bases functions. This,
however, can be avoided by recognizing that in the ${\rm C_n}$ group, states 
with wavevectors 'k' and '-k' are degenerate. We can therefore construct a 
linear combination of the 'k' 
and '-k' states which is real. The symmetry representations in the ${\rm C_n}$ 
group would then correspond to the labels $A$, $B$ and $E$, with the characters
in the $E$ representation given by ${\rm 2\cos}(rk)$ under the symmetry 
operation ${\rm C^r_n}$, with ${\rm k=\pi / n}$. The parity operation commutes 
with the spatial symmetry operations and the full point group of the system 
would then correspond to the direct product of the two groups. Since both 
parity and spatial symmetries can be easily incorporated in a constant ${\rm 
M_S}$ basis, we do not encounter the difficulties endemic to the VB theory.

The generation of the complete basis in a given Hilbert space requires a simple
representation of a state on the computer. This is achieved by associating
with every state a unique integer. In this integer, we associate ${\rm n_i}$ 
bits with spin ${\rm s_i}$, such that ${\rm n_i}$ is the smallest integer for 
which ${\rm 2^{n_i} \ge (2s_i+1)}$. In the integer that represents the state 
of the cluster, we ensure that these ${\rm n_i}$ bits do not take values which 
lead to the ${\rm n_i}$ bit integer value exceeding ${\rm (2s_i+1)}$. For each 
of the allowed bit states of the ${\rm n_i}$ bit integer, we associate an 
${\rm M_{S_i}}$ value between ${\rm -s_i {\rm ~and}~ s_i}$. For a spin cluster 
of 'n' spins, we scan all integers of bit length ${\rm N=\overset{n}{\underset
{i=1}{\Sigma}}n_i}$ and verify if it represents a basis state with the desired 
${\rm M_S}$ value. In Fig. 1, we show a few basis functions with specified 
${\rm M_s}$ value for some typical ferric wheels along with their bit 
representations and the corresponding integers. Generation of the bases states 
is usually a very fast step, computationally. Generating the basis as an ordered 
sequence of integers that represent them also allows for rapid generation of 
the Hamiltonian matrix elements as will be seen later.  

Symmetrization of the basis by incorporating parity and spatial symmetries
involves operation on the constant ${\rm M_S=0}$ basis by the symmetry 
operators. Since spatial symmetry operators permute the positions of 
equivalent spins, every spatial symmetry operator operating on a basis 
function generates another basis function. Every symmetry operator can be 
represented by a correspondence vector whose ${\rm i^{ th}}$ entry gives 
the state that results from operating on the ${\rm i^{th}}$ state by the 
chosen operator. This is also true for the parity operator, in the ${\rm 
M_S~=~0}$ subspace. 

The first step in constructing symmetry adapted linear combinations is to 
represent the symmetry operators in the chosen basis as matrices. In our
case, the symmetry operators are such that symmetry operation by any 
operator on a basis state leads to a resultant which is a single basis
state. Thus all our symmetry operators can be represented as vectors; the
entry in position 'i' gives the index of the basis function generated by
symmetry operation on the basis state 'i'. Since the basis is very large,
it is prohibitive to store and manipulate the full basis together with
all the associated symmetry vectors.  To avoid these difficulties, we have 
constructed the symmetry matrices in small invariant subspaces. These 
invariant subspaces are obtained by sequentially choosing a state and 
operating on it by all the symmetry operators. This gives rise to a set 
of states on which we again operate by all the symmetry operators, and 
continue this process until no new basis states are generated. The 
collection of all these basis states resulting at the end of this process 
is the invariant subspace. We can set-up symmetry combinations of the basis 
states in each of the invariant subspaces independently. The symmetry 
combinations can now be obtained operating on each state by the group 
theoretic projection operator,
\begin{equation}
 {\hat P}_{\Gamma_i}~=~ {{1} \over {h}}~ \underset{R}{\Sigma}~\chi_{\Gamma_i}(R)
{\hat R}
\end{equation}
on each of the basis states of the invariant subspace. Here $\Gamma_i$ is 
the $i^{th}$ irreducible representation, ${\hat R}$ is the symmetry operation 
of the group and $\chi_{\Gamma_i}(R)$ is the character under ${\hat R}$ in 
the irreducible representation $\Gamma_i$. This process is repeated with
the next basis state that has not appeared in any of the invariant subspaces
already constructed. The process comes to an end when all the basis states 
have appeared in any one of the invariant subspaces.

The resulting symmetrized basis is usually over complete in each of the
invariant subspaces. The linear dependencies can be eliminated by a 
Gram-Schmidt orthonormalization procedure. However, in most cases, ensuring 
that a given basis function does not appear more than once in a symmetrized 
basis is sufficient to guarantee linear independence and weed out the 
linearly dependent states. A good check on the procedure is to ensure that 
the dimensionality of the symmetrized space in the invariant subspace agrees 
with that calculated from the traces of the reducible representation 
obtained from the matrices corresponding to the symmetry operators in 
the chosen invariant subspace. Besides, the sum of the dimensionalities of 
the symmetrized spaces should correspond to the dimensionality of the 
unsymmetrized invariant subspace in each of these subspaces.

Generation of the Hamiltonian matrix is rather straight forward and involves
operation of the Hamiltonian operator on the symmetry adapted basis. This
results in the matrix ${\mathbf SH}$, where ${\mathbf S}$ is the symmetrization
matrix representing the operator ${\hat P}_{\Gamma_i}$ and ${\mathbf H}$ is
the matrix whose elements $h_{ij}$ are defined by
\begin{equation}
{\rm
{\hat H} |i> = \underset{j}{\Sigma}~ h_{ij} |j>.
}
\end{equation}
The states \{i\} correspond to the unsymmetrized basis functions. The 
Hamiltonian matrix in the symmetrized basis is obtained by right multiplying
the matrix ${\mathbf SH}$ by ${\mathbf S^\dagger}$.  The resulting symmetric
Hamiltonian matrix is stored in the sparse matrix form and the matrix
eigenvalue problem is solved using the Davidson algorithm \cite{davidson}.

Computation of the properties is easily done by transforming the eigenstate
in the symmetrized basis into that in the unsymmetrized basis. Since the
operation by any combination of spin operators on the unsymmetrized basis
can be carried out, all relevant static properties in different eigenstates
can be obtained in a straightforward manner.

\subsection{Quantum Spin Dynamics} 

We have studied the dynamics in ferric wheels by setting up the Hamiltonian 
matrix in the desired ${\rm M_S}$ space, which in all cases is restricted to 
${\rm M_S =0, 1}$ and $2$. In each subspace we have obtained a few low-lying 
states using the Davidson algorithm \cite{davidson}. We have also calculated 
the spin-spin correlation functions in each of the states. Using the spin-spin 
correlation functions, we have computed the expectation value of ${\rm S^2_
{total}}$ operator, from which we have identified the total spin of the state. 
We observe from the eigenvalue spectrum of all the ferric wheels that ground 
state and the first excited state are spin singlet (S=0) and triplet (S=1) 
state respectively, they belong to two different spatial symmetry subspaces. 
So they will not mix unless there is a perturbation which spoils the ${\rm C_
{10}}$ symmetry of the molecule. We also notice that the first (triplet) 
and the second (quintet) excited states again fall into different symmetry 
subspaces. This is true in all the ferric wheels we have studied.

To study quantum dynamics we have considered the following Hamiltonian 
\cite{zvezdin},
\begin{equation}
{\rm 
\hat{H} = E_s - D~\hat{S}_{z,total}^2 + c~(\hat{S}_{x,total}^2 -
\hat{S}_{y,total}^2) - g~h(t)~\hat{S}_{z,total} 
}
\label{ham2}
\end{equation}
Here D is the quadratic anisotropy factor, $g$ is the Land\'e $g$-factors for
the iron(III) spin respectively, and h(t) is the time-dependent magnetic field, 
expressed as ${\rm h(t) = H_0 \cos(\omega t)}$, where $\omega$ is the frequency 
at which field is ramped and ${\rm H_0}$ is the amplitude of the field. We have 
chosen D = $8.8 \times 10^{-3}$ and c = $10^{-3}$ (in units of J) in accordance 
with the experimental values for ${\rm Fe_{10}}$. We take $g = 2.0$. The 
constants ${\rm E_s}$ in Eq. (\ref{ham2}) correspond to the lowest energies 
obtained from Eq. (\ref{ham1}). The second order anisotropy term allows 
transitions between states with ${\rm \Delta M_S = \pm 2}$. Both second and 
third terms in Eq. (\ref{ham2}) arises due to magnetocrystalline anisotropy. 
The exact form of the anisotropy in ferric wheel is not very well established. 
We have included anisotropy terms only up to second order in the spin variables. 
The anisotropy in the plane can be formed artificially, e.g. by means of 
external electric or magnetic fields, pressure or using anisotropic substrate 
\cite{zvezdin}.

The Hamiltonian in Eq. (\ref{ham2}) does not allow the mixing of the (singlet)
ground state, (triplet) first excited state and (quintet) second excited state 
in ${\rm Fe_{10}}$ because of symmetry reason already mentioned earlier. To 
observe spin dynamics we apply a constant magnetic field that leads to triplet 
being the ground state of the system. This constant shifting field plays 
here the role of a chemical potential and is used to provide the optimal 
conditions for observing the considered quantum spin dynamics. We can represent
the external field as 
\begin{equation}
{\rm 
B ~=~ B_0 ~+~ H_0 \cos (\omega t)
}
\end{equation}
To study the evolution of magnetization as a function of applied oscillatory 
field we start from a fixed axial field amplitude (in units of ${\rm J/ \hbar}
$) and time is increased in small time step of $\Delta$t =0.01 (${\rm \hbar 
/J}$), frequency of the field is kept constant. 

We have studied the time evolution of the system by solving the time-dependent
Schr\"odinger equation,
\begin{equation}
{\rm 
i\hbar \frac{d\psi}{dt} = \hat{H}(t) \psi ~.
}
\end{equation}
We assume the system to be in the state with $S=1,~M_S=-1$ at time $t=0$. This 
is the initial state which is time evolved according to the equation
\begin{equation}
\psi(t+\Delta t) = e^{-i \hat{H}(t+\frac{\Delta t}{2}) \Delta t / \hbar} ~
\psi(t)~.
\end{equation}
The evolution is carried out by explicit diagonalization of the Hamiltonian
matrix ${\bf H}(t+\frac{\Delta t}{2})$, and using the resulting eigenvalues
and eigenvectors to evaluate the matrix of the time evolution operator $e^{-i
\hat{H}(t+\frac{\Delta t}{2}) \Delta t/\hbar}$. We set up the Hamiltonian 
matrix for time evolution in the truncated basis of 3 states corresponding to 
total spin $S$=1. We repeatedly carry out the the time evolution in small time 
steps of size $\Delta t$ to obtain time evolution over longer periods.

\section{Results and Discussion}
\subsection{Analysis of Low-lying Spectrum}

We have solved the exchange Hamiltonian (Eq. (\ref{ham1})) exactly using the
method mentioned earlier to obtain the low-lying eigenvalue spectrum for 6, 8, 
10 and 12 site iron(III) rings. We find that ground state, first, second and 
third excited states are respectively singlet, triplet, quintet and heptet 
for all ferric wheels. We notice that there is no accidental degeneracy 
between the 
energy levels belonging to different symmetry subspaces. Ground state switches 
between A$^+$ (k =0) and B$^+$ (k = $\pi$) subspace for even and odd value of 
'm' respectively, in N=2m ring systems. This was also observed by Mattheiss in 
case of a spin-1/2 chain and can be understood from Marshall's sign rule. 
The gap between the ground state and the first excited state is shown in
Fig. 2 as a function of inverse ring size. According to Haldane conjecture,
the gap should extrapolate to zero. The extrapolated value, while small is 
still finite, suggesting that in these rings the finite size effects are
still at play at the ring sizes we have studied. 

Using the exchange constants estimated for the different ring systems, we 
estimate the gap between the ground state and the first excited state to be 
22.67 K, 11.81 K, and 6.88 K for ${\rm Na : Fe_6}$, ${\rm Fe_{8}}$, 
${\rm Fe_{10}}$ respectively. Our calculated values compare very well with 
the experimental values, which are 22.0 K, 12.1 K and 6.45 K for ${\rm Na : 
Fe_6}$, ${\rm Fe_{8}}$ and ${\rm Fe_{10}}$ respectively \cite{gatteschi2}. 
This agreement shows that for all practical purposes ferric wheels can safely 
be assumed as rings neglecting the slight deviation from the exact circular 
geometry. Our calculated gap for and ${\rm Fe_{12}}$ is 12.09 K corresponding 
to the exchange constants predicted from experiments. However, experimental 
estimate of this gap is lacking. 

If we define $\delta_i$ to be the energy difference between the $i$-th 
excited state and the ground state, we find from Fig. 3 that the following 
relationship is satisfied for the ferric wheels, 
\begin{equation}
\delta_i ~=~ {\frac{S_i(S_i~+~1)}{2}} \times E_1
\end{equation}
where E$_1$ is the energy gap between the ground state and the first excited 
state. This indicates that the lowest spin state obeys the Lande interval rule,
in agreement with Taft's conjecture. If we assume $E_1$ to be the inverse 
of the moment of inertia, then the above expression gives the rotational
energy of a rigid rotor in a state with quantum number $S_i$. Thus, the 
spin states of ferric wheels can be viewed as the quantized states of a
rigid rotor.

In Fig. 4 we have shown the dispersion spectrum of ferric wheels. The value
of $k$ corresponding to a wavefunction $\psi$ can be defined as, 
\begin{equation}
T \psi ~=~ e^{ik} \psi
\end{equation}
where $T$ is the translation operator which in case of a ring rotates the ring
by one lattice spacing. We have used the spatial $C_n$ symmetry of the ferric
wheels, which enables us to identify the value of $k$ easily for a specified
eigenfunction. We find that, in the momentum ($k$) sector which contains the 
ground state, the lowest excitation is to a quintet state. For all the other 
$k$-sectors the lowest excitation is to a triplet state. Previous studies 
\cite{pearson} on antiferromagnetic spin-1/2 Heisenberg chains show that the
excitation spectrum is given by $\hbar \omega ~=~ (\pi/2) J |\sin k|$, 
where $k$ 
is the wave vector of the excited states measured with respect to that of the 
ground state. The simple antiferromagnetic spin-wave theory based on the use of
the Holstein-Primakoff transformation for each sublattice, leads to excitation
spectrum ($S$ is the magnitude of the individual spin $S_i$), 
\begin{equation}
\hbar \omega ~=~ 2~J~S~|\sin k|.
\label{spw}
\end{equation}
This relation is supposed to be correct for $S \rightarrow \infty$. We notice 
that the excitation spectrum for ferric wheels can be fitted to a $|\sin k|$ 
kind of function. Data points in $k$=0 or $k=\pi$ diverges from the above 
sinusoidal function. This may be a finite size effect. In the thermodynamic 
limit of infinite chain length excitation spectrum of ferric wheel will be 
similar to Eq. (\ref{spw}).

We have also calculated the spin-spin correlation function ($\sum_{ij}<S_i^z
S_j^z>$) of ferric wheels and Fourier transformed it to find out the structure 
factor ($S(q)$). In Fig 5.(a - d) we have plotted the structure factor as a 
function of wave vector $q$ for different symmetry subspaces. In all the cases 
$S(q)$ shows a peak at $q~=~\pi$. The ground state is a Ne\'el ordered state. 
It also signifies that ground state will undergo a spontaneous distortion with
wave vector $\pi$. We find that the weight of $S(q)$ is more in case of ground 
state than excited states.

\subsection{Evolution of Magnetization in presence of an {\it ac} field}

We have studied the evolution of magnetization in presence of an axial {\it ac}
magnetic field varying the amplitude of the field. We have kept the frequency
of the field fixed at $\omega$=10$^{-3}$. We calculate the magnetization at
each time step. When we draw a smooth curve for the time evolution over long
time periods, we find a sinusoidal motion,
\begin{equation}
M(t) \sim \cos(\Omega t)
\end{equation}
which can be seen in Fig. 6. Unexpectedly the frequency $\Omega$ of this 
sinusoidal motion does not correspond to an eigenfrequency of the system or
to the period of the external field. When we change the amplitude H$_0$ of the
field, the period of the magnetization changes, which is shown in Fig 6. 
We find a regular dependence of $\Omega$ as a function of H$_0$, as in Fig 7. 
Frequency of oscillation $\Omega$ becomes very small for some values of H$_0$,
again it increases but to a lower value than the previous $\Omega_{max}$.
$\Omega_{max}$ values can be fitted to a exponentially decaying curve.
This nontrivial resonance can be understood from the viewpoint of Floquet's
theorem \cite{floquet}. Nonadiabatic transitions are possible whenever system 
becomes metastable. This resonance can be observed if two states with different
magnetizations are nearly degenerate leading to large tunneling cross sections
between the states resulting in oscillation of the magnetization. This 
magnetization oscillation can be related to that of macroscopic quantum 
coherence, which is predicted by Zvezdin {\it et. al.} \cite{zvezdin}. 

\subsection{Torque Magnetometry}

Cornia et. al. have used a novel cantilever torque magnetometry technique to 
study the spin-state crossover in ferric wheels. The torque {\bf T} experienced 
by a magnetically anisotropic substance in a uniform magnetic field {\bf B}
is given by 
\begin{equation}
{\mathbf T} = {\mathbf M} \times {\mathbf B}
\end{equation}
where {\bf M} is the magnetization of the sample. {\bf T} vanishes when the 
magnetic field is applied along one of the principle directions ${\hat x}$,
${\hat y}$, ${\hat z}$ of the susceptibility tensor, since in this case {\bf M} 
and {\bf B} are collinear. The ${\hat y}$ component of the torque operator 
{\bf T}$_y$ can be easily obtained for an applied magnetic field of the form
\begin{equation}
{\mathbf B} = B ( \cos \theta {\hat z} + \sin \theta {\hat x}),
\end{equation}
in Eq. 5, and is given by,
\begin{equation}
<{\mathbf T}_y> = -g \mu_B B (<S_x> \cos \theta - <S_z> \sin \theta)
\end{equation}
where $<S_\alpha>$ = $\sum_{i=1}^N$ $<S_{i,\alpha}>$ is the ground state 
expectation value of the component $\alpha$ of the total spin operator.
We have computed $<{\mathbf T}_y>$ for Fe$_{12}$ on the basis of the 
eigenvectors of the Hamiltonian in Eq. 5. When $\theta$ is ${\frac{\pi}{4}}$
then torque is maximal and $<{\mathbf T}_y>$ is just the difference between
the ${\hat x}$ and ${\hat z}$ spin components. In Fig. 8 we have shown the 
variation of $<{\mathbf T}_y>$ with magnetic field. We can clearly observe 
the step behavior of the torque component which is a manifestation of the 
level crossing of singlet,triplet and quintet states.

\section{Summary and Outlook}
We have implemented a general and efficient procedure that allows us to block
factorize the spin Hamiltonian matrix based on its invariance under cyclic
symmetry and parity operation. This method can be used in general for systems
of other symmetries also. We have obtained the low-lying eigenvalue spectrum 
of ferric wheels up to ${\rm Fe_{12}}$ using the C$_n$ rotational symmetry of 
the molecules. We have also analyzed the dispersion spectrum and structure 
factor.  To reproduce the low temperature properties of ferric wheels
we need to know the low-lying eigenvalue spectrum of these systems. 
We have also studied the dynamics of ferric wheel evolving a proper initial 
state. We observe nontrivial oscillation of magnetization in presence of an 
alternating magnetic field. We have also obtained the torque of the ferric
wheels in the presence of a nonaxial magnetic field and find that the torque
also exhibits step like behavior with field. Evidently a study including the 
effect of nonzero temperature on this oscillation is a challenging problem of 
future research.

{\bf Acknowledgement:} We thank CSIR, India and DAE-BRNS, India for financial 
support.

\newpage

\noindent{\bf Figure Captions}
\vspace{0.5cm}

\noindent {1.} Representative ${\rm M_s=0}$ state in (a) 6 spin-$\frac{1}{2}$
cluster (b) ${\rm Fe_{12} }$ wheel with all the sites having spin
 $\rm {S=\frac{5}{2}}$. Numbers in paranthesis correspond to the ${\rm M_s}$ 
value at the site. Bit representation as well as the integer value are given 
just below the diagrams.

\noindent {2.} Plot of ground state energy (in unit of J) vs. inverse system 
size for ferric wheels up to Fe$_{12}$. 

\noindent {3.} Plot of lowest energy (in unit of J) in every total spin sector 
vs. the total spin (S$_{tot}$) in case of Fe$_{12}$. It clearly obeys Taft's 
conjecture.

\noindent {4.} Plot of energy (in unit of J) vs. momentum vector $k$ for 
${\rm Fe_6}$, ${\rm Fe_8}$, ${\rm Fe_{10}}$ and ${\rm Fe_{12}}$.

\noindent {5.} (a) Plot of static structure factor (S$_q$ (in arbitary units))
for ${\rm Fe_6}$ in all the 
subspaces. Here ground state lies in B$^+$ subspace. In (b), (c) and (d) we
have plotted the static structure factor in A and B subspaces for ${\rm Fe_8}$,
${\rm Fe_{10}}$ and ${\rm Fe_{12}}$ respectively. In the doubly degenerate
subspaces the weight of S(q) is much smaller than that in the ground state.

\noindent {6.} Plot of evolution of magnetization in presence of an alternating
axial magnetic field of three different amplitudes (H$_0$(in unit of J/$\hbar$)).   

\noindent {7.} Field amplitude (H$_0$) dependence of frequency $\Omega$ 
(defined in the text). Dotted line shows the exponential nature of 
$\Omega_{max}$.

\noindent {8.} Plot of $\hat y$ component of torque (in unit of ${\frac{J}
{\hbar-rad}}$)with change in magnetic field applied at $\theta = \pi/4$.

\begin{figure}[hp]
\centerline{\includegraphics[height=12cm ]
{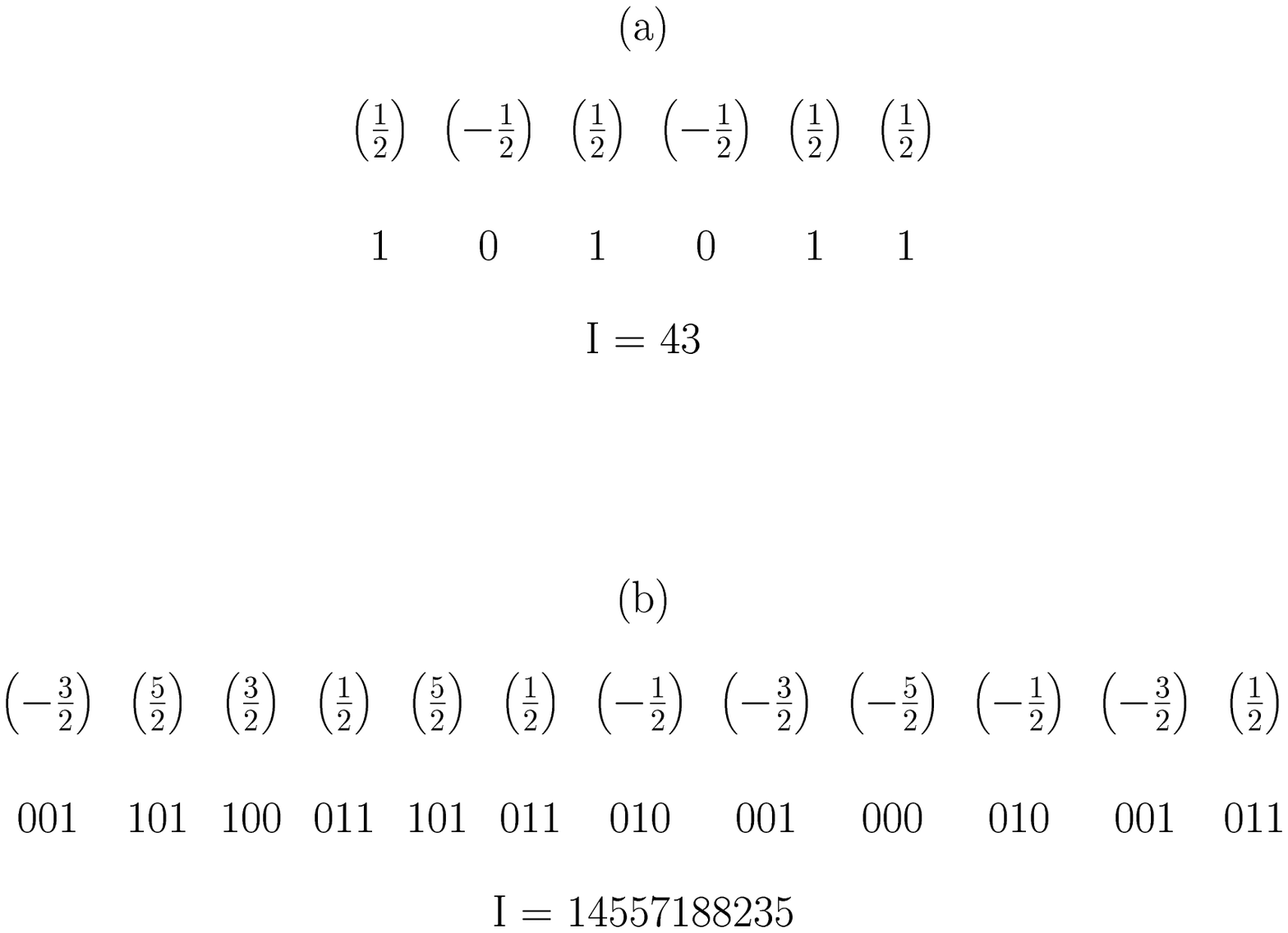}}
\end{figure}

\begin{figure}[hp]
\centerline{\includegraphics[height=18cm,width=15cm ]
{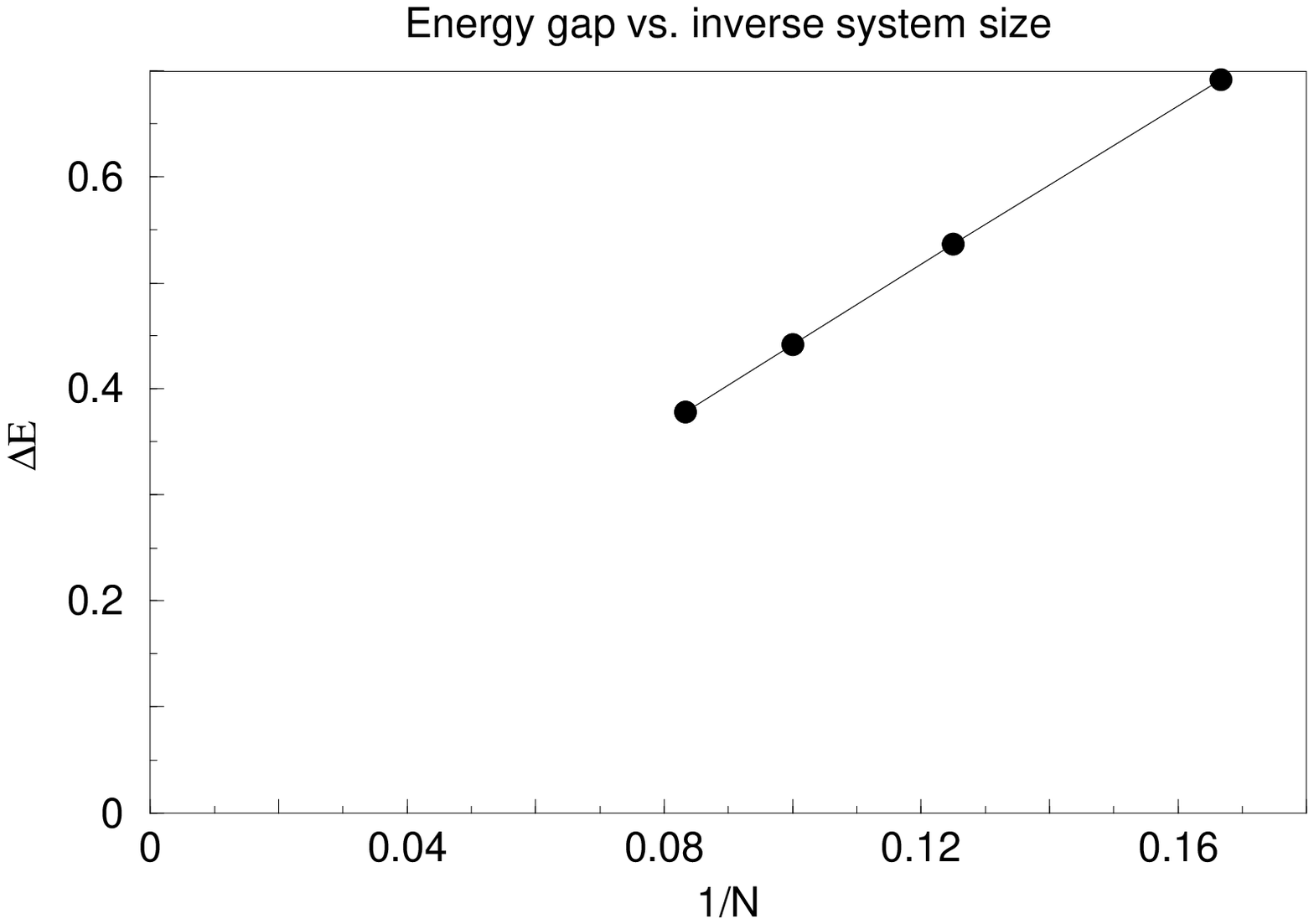}}
\centerline{\Large{Fig. 2 }}
\end{figure}

\begin{figure}[hp]
\centerline{\includegraphics[height=18cm,width=15cm ]
{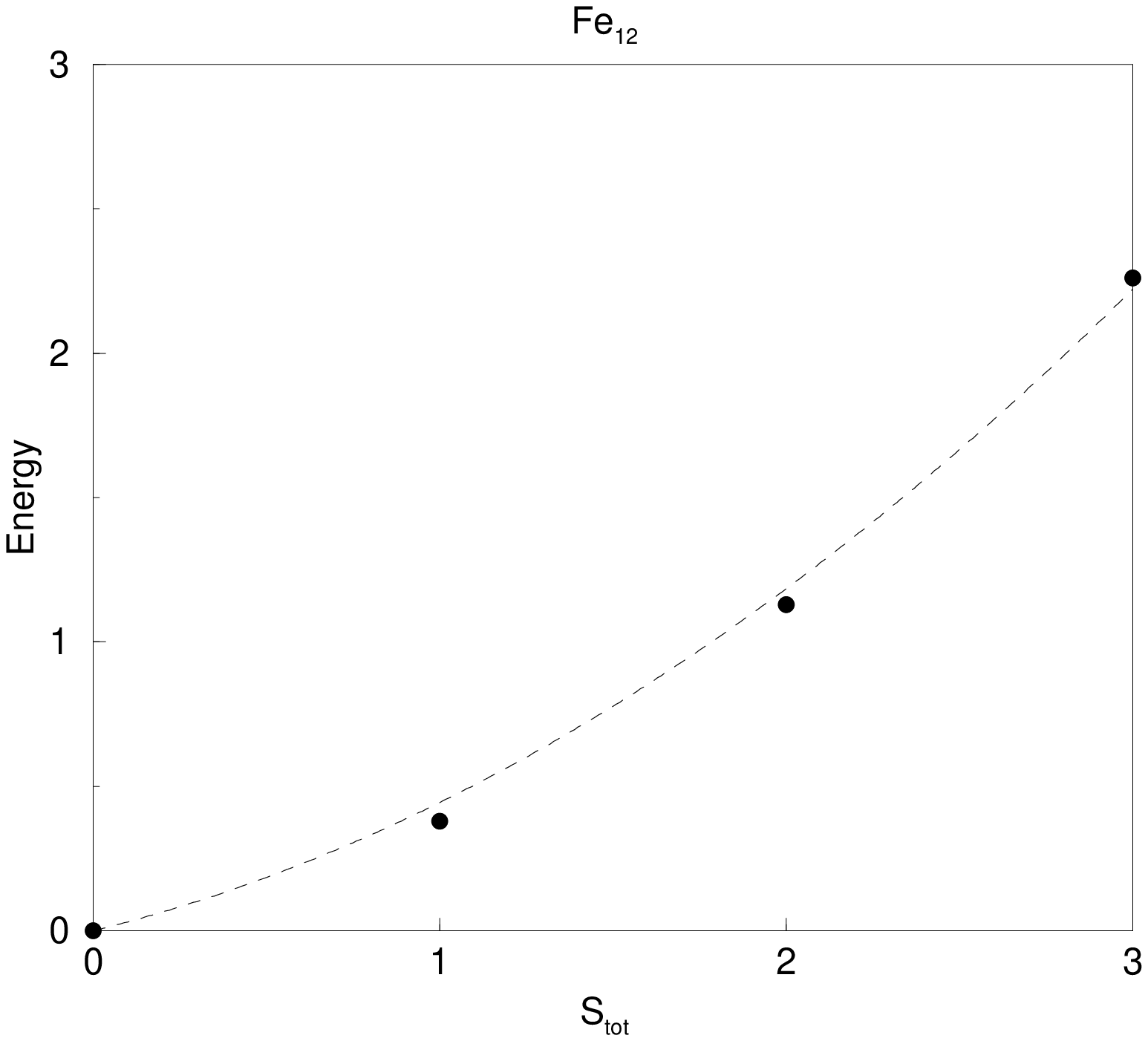}}
\centerline{\Large{Fig. 3 }}
\end{figure}

\begin{figure}[hp]
\centerline{\includegraphics[height=18cm,width=15cm ]
{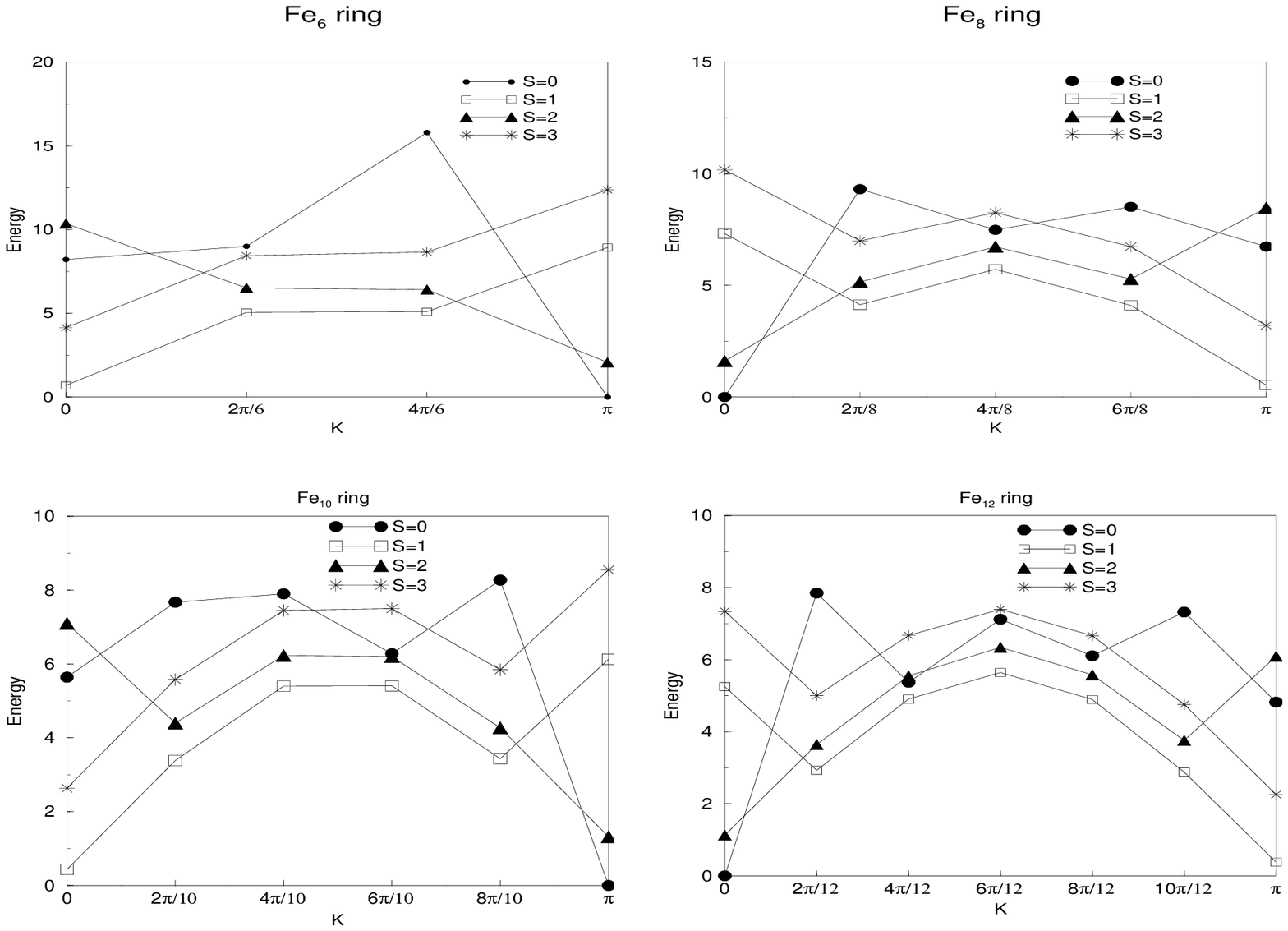}}
\centerline{\Large{Fig. 4}}
\end{figure}

\begin{figure}[hp]
\centerline{\includegraphics[height=18cm,width=15cm ]
{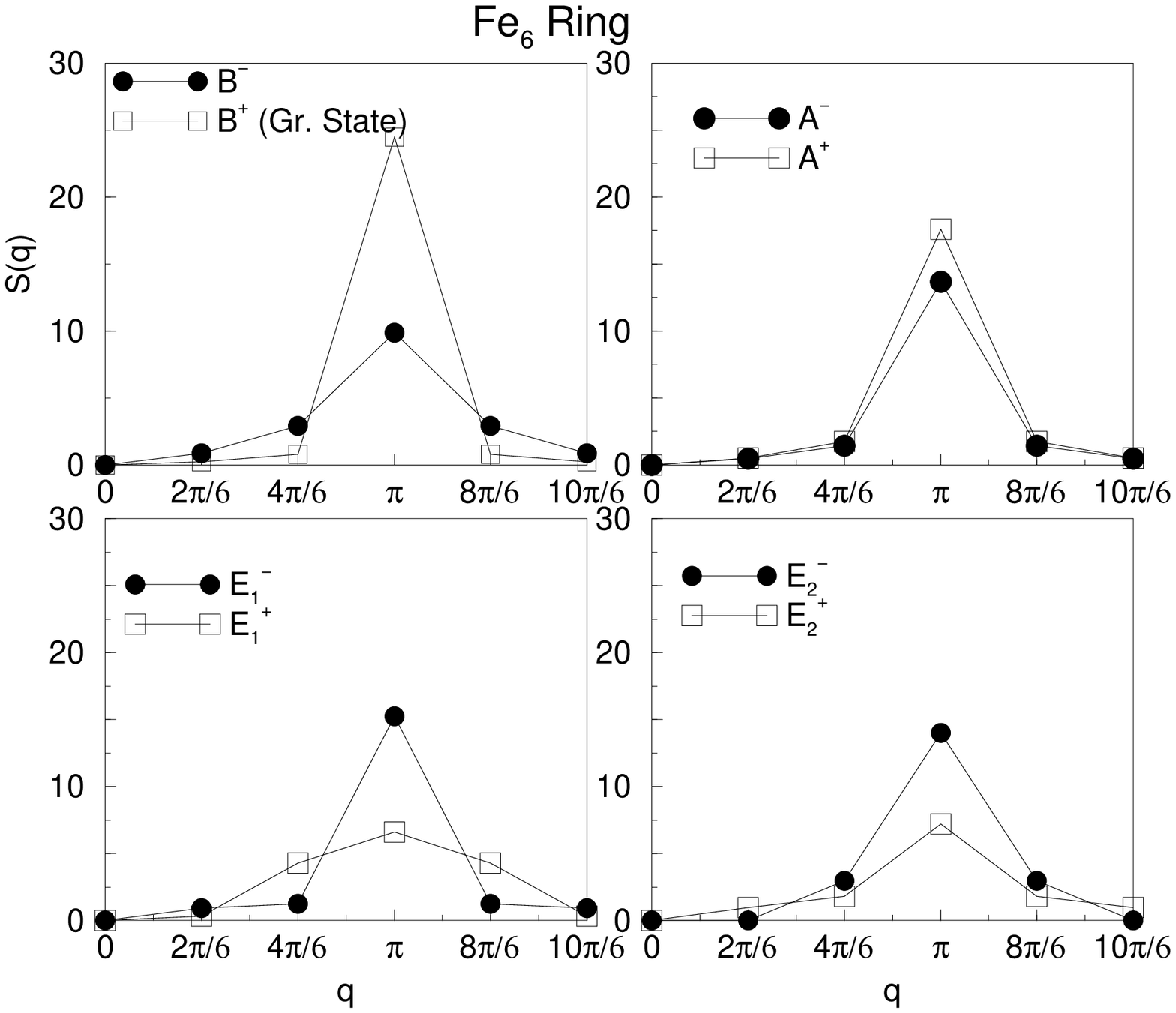}}
\centerline{\Large{Fig. 5 (a) }}
\end{figure}

\begin{figure}[hp]
\centerline{\includegraphics[height=18cm,width=15cm ]
{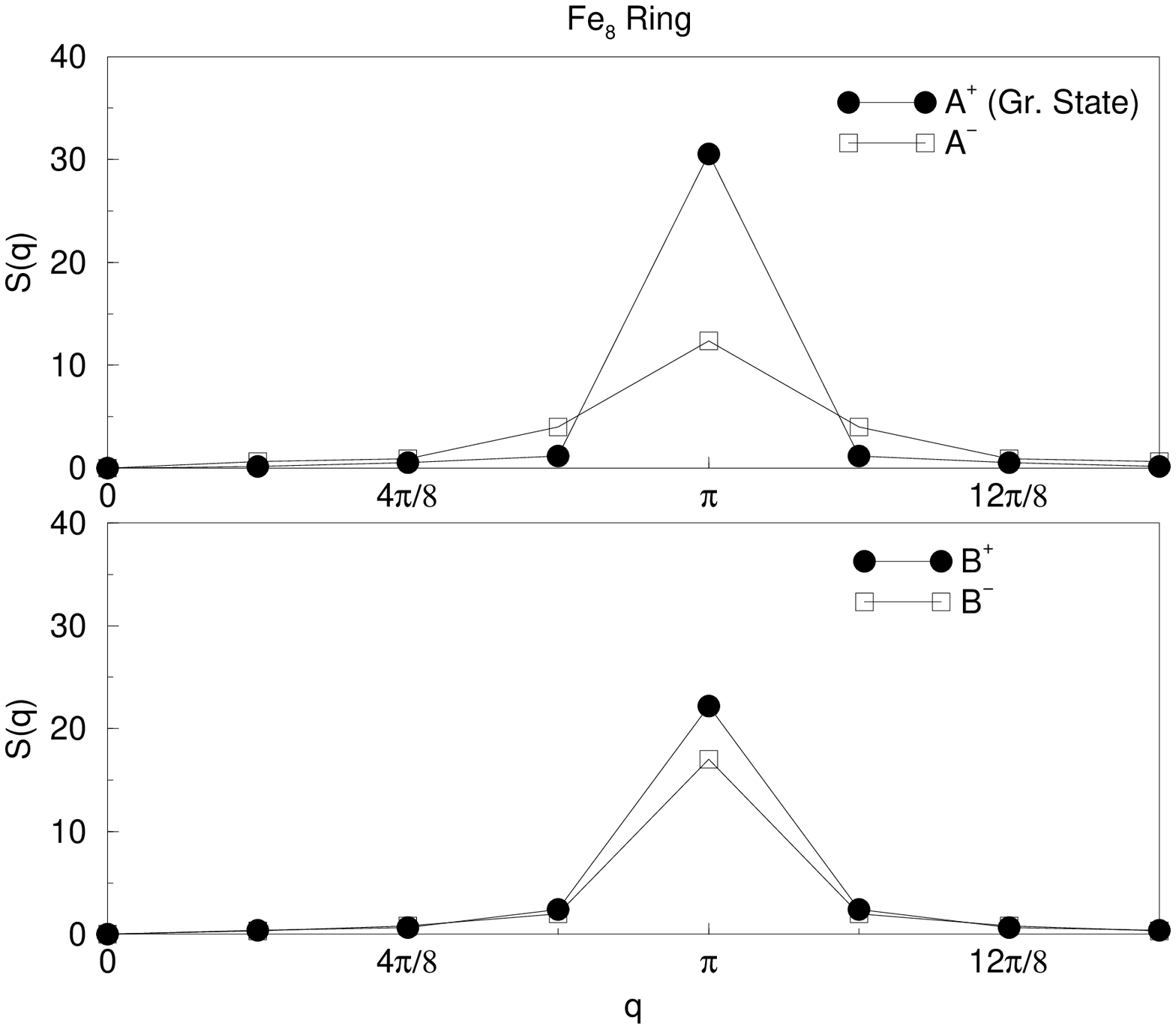}}
\centerline{\Large{Fig. 5 (b) }}
\end{figure}

\begin{figure}[hp]
\centerline{\includegraphics[height=18cm,width=15cm ]
{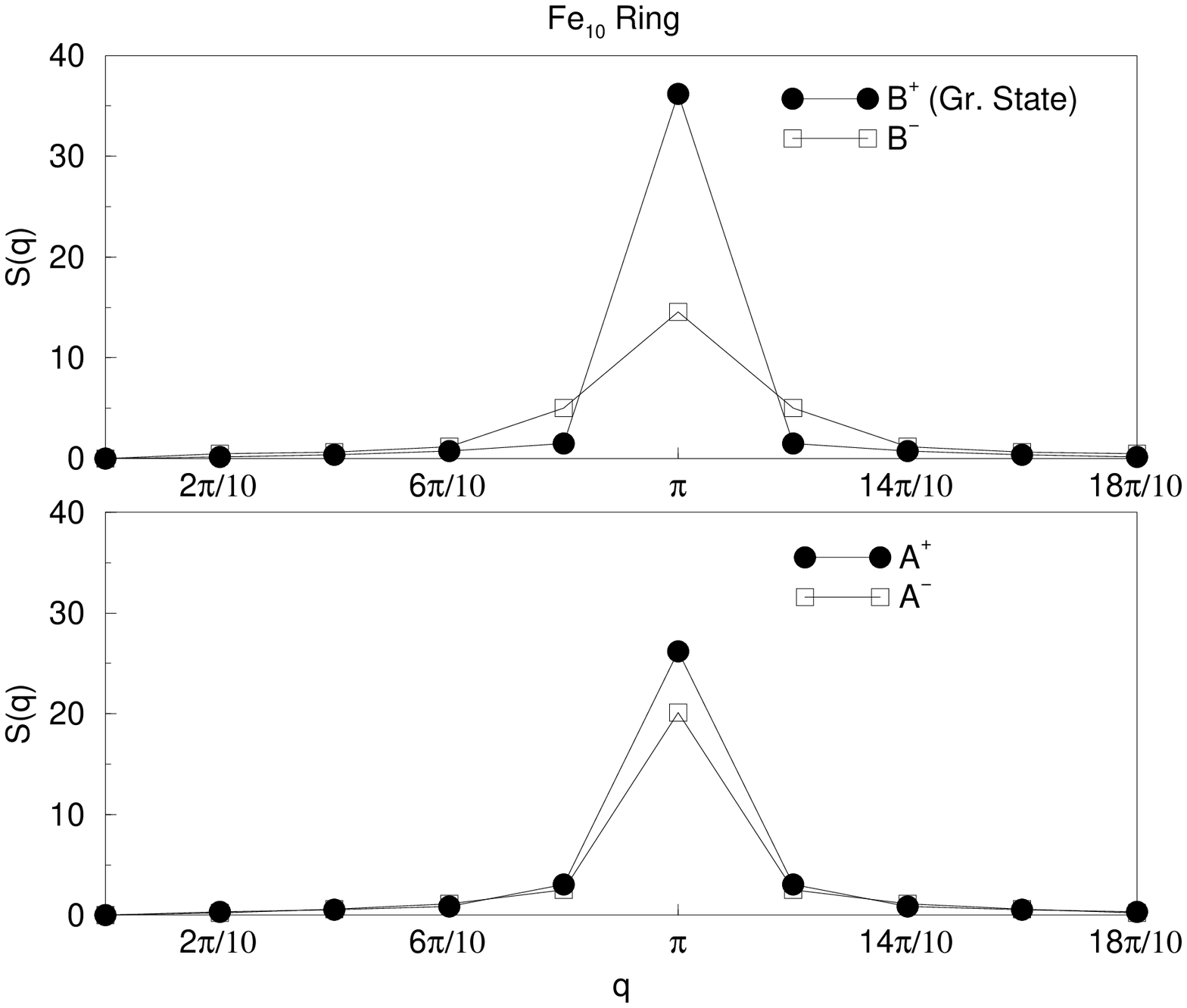}}
\centerline{\Large{Fig. 5 (c) }}
\end{figure}

\begin{figure}[hp]
\centerline{\includegraphics[height=18cm,width=15cm ]
{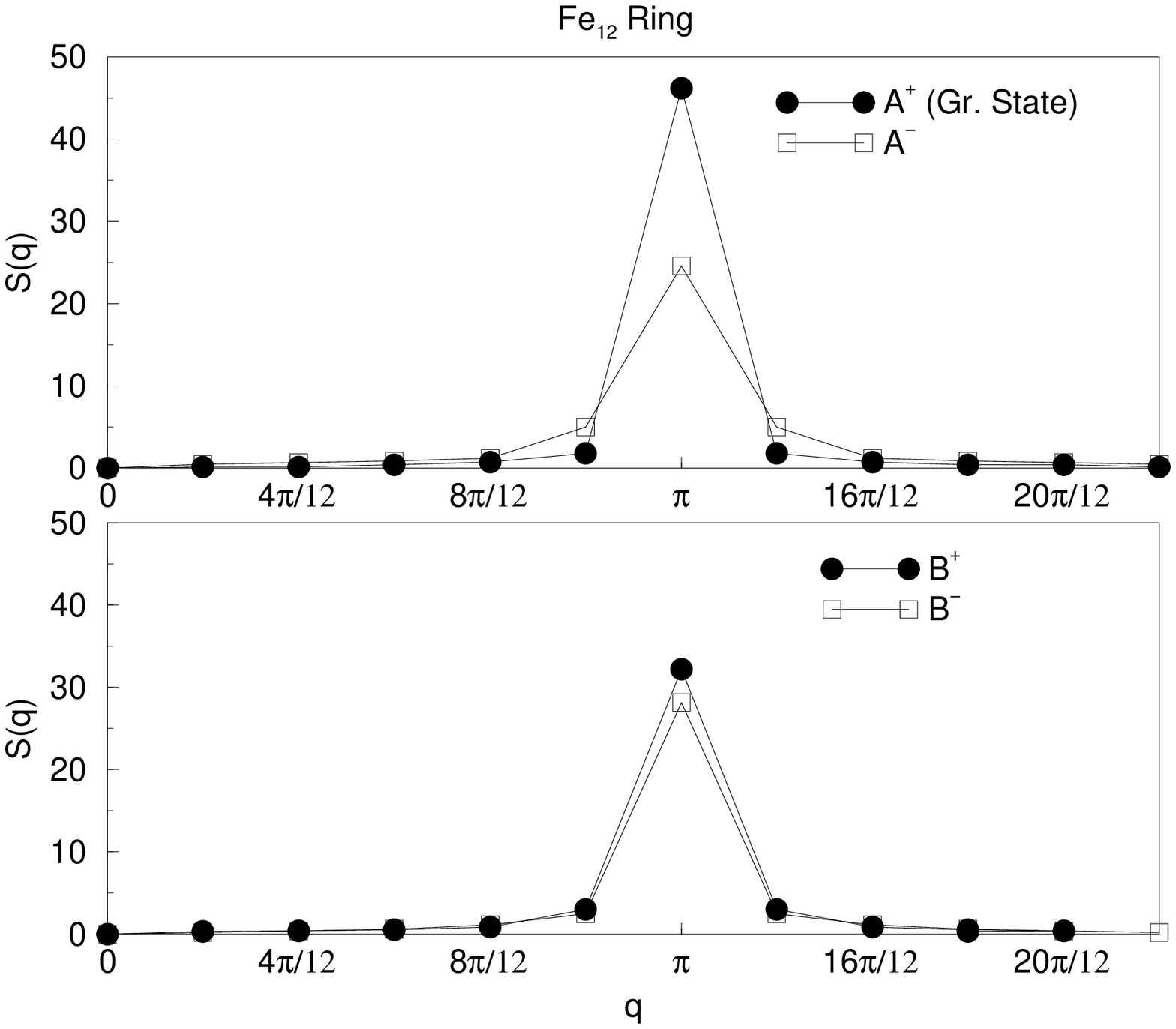}}
\centerline{\Large{Fig. 5 (d) }}
\end{figure}

\begin{figure}[hp]
\centerline{\includegraphics[height=18cm,width=15cm ]
{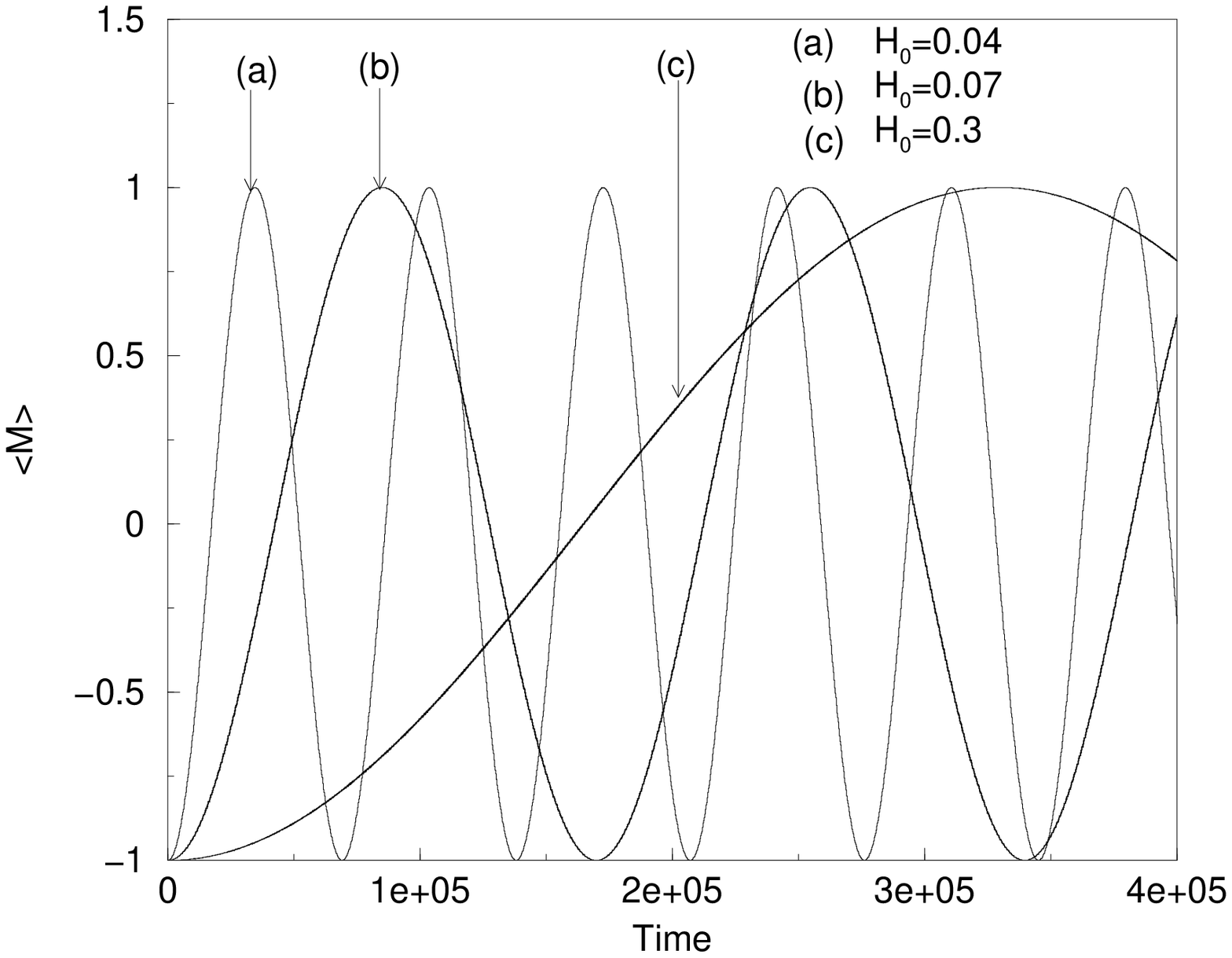}}
\centerline{\Large{Fig. 6 }}
\end{figure}

\begin{figure}[hp]
\centerline{\includegraphics[height=18cm,width=15cm ]
{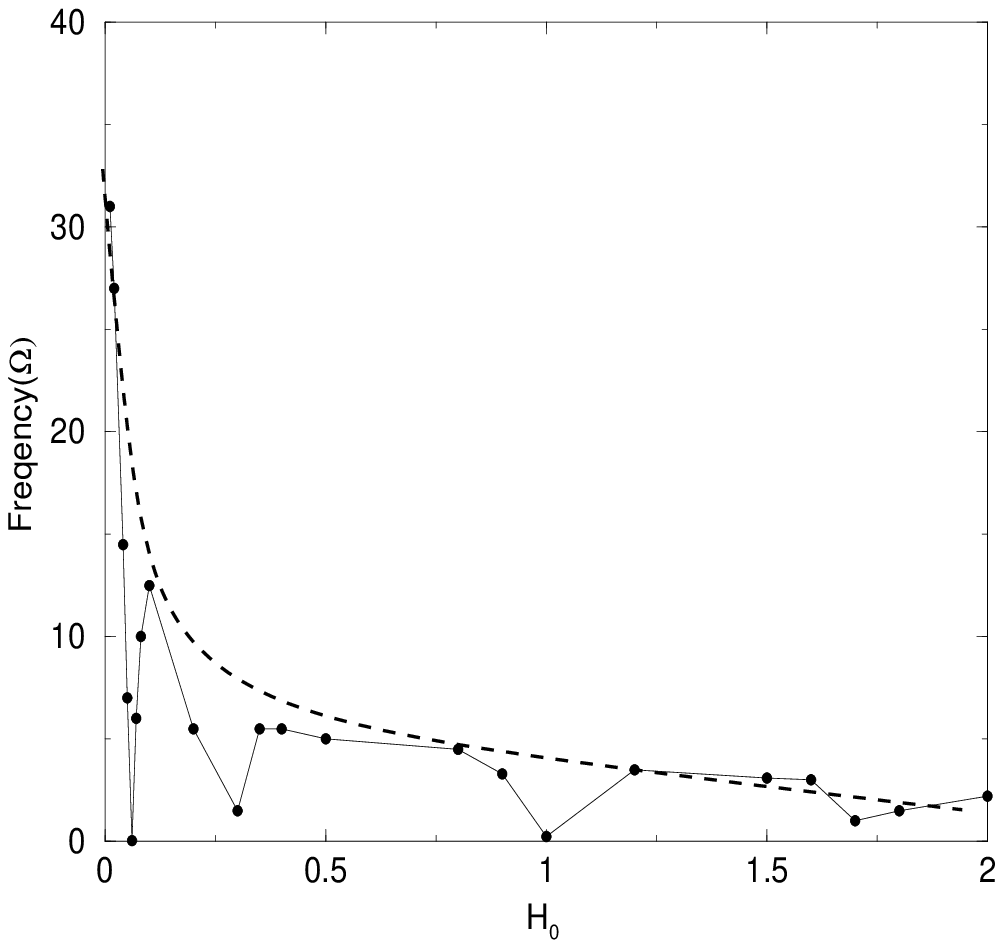}}
\centerline{\Large{Fig. 7 }}
\end{figure}

\begin{figure}[hp]
\centerline{\includegraphics[height=18cm,width=15cm ]
{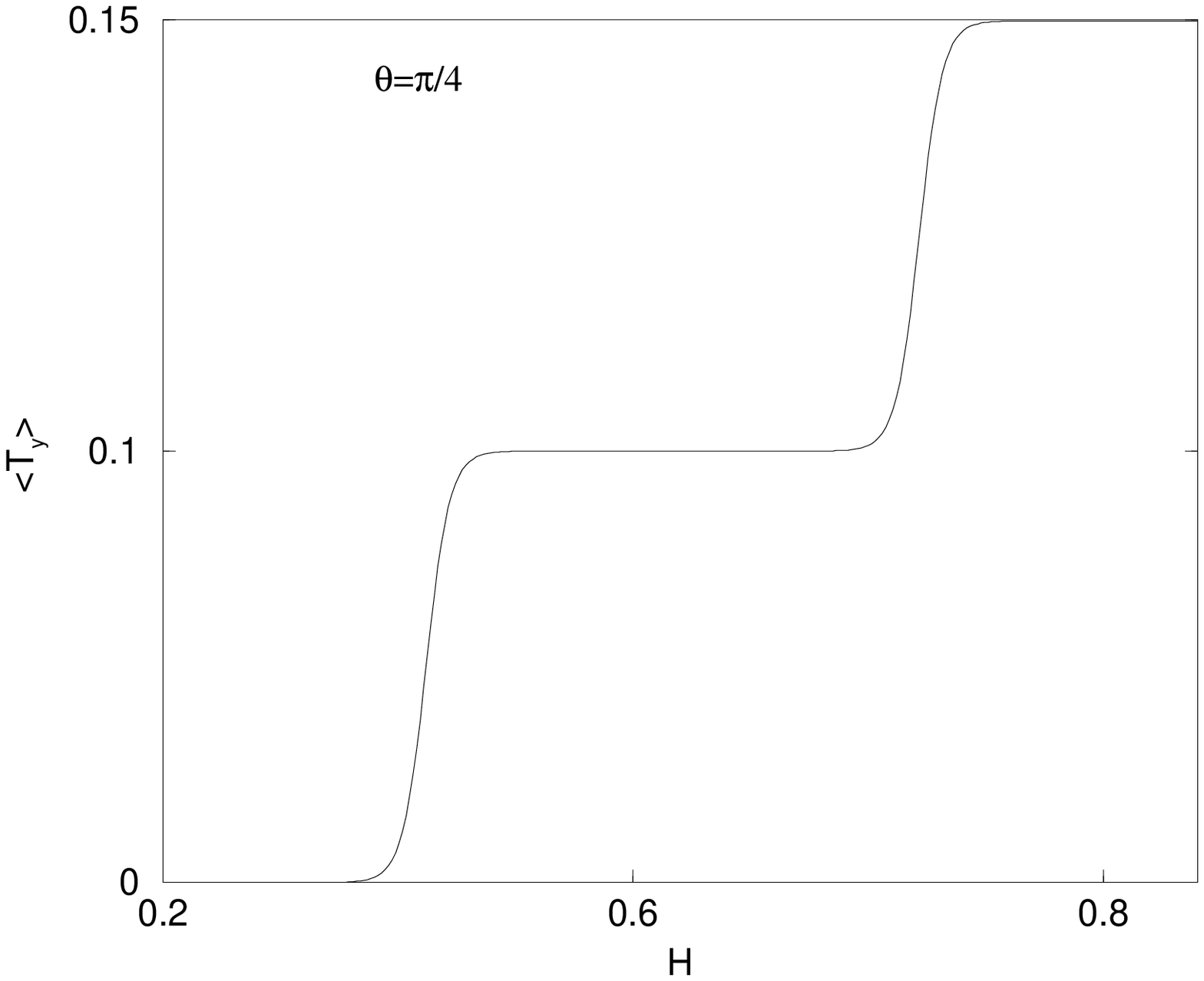}}
\centerline{\Large{Fig. 8 }}
\end{figure}


\begin{thebibliography}{99}

\bibitem{lippard} K. L. Taft, C. D. Delfs, G. C. Papaefthymiou, S. Foner, 
D. Gatteschi and S. J. Lippard, J. Am. 
Chem. Soc. {\bf 116}, 823 (1994).

\bibitem{synth} G. L. Abbati, A. Cornia, A. C. Fabretti and W. Malavasi,
Inorg. Chem. {\bf 36}, 6443 (1997) ; A. Caneschi, A. Cornia and S. J. Lippard,
 Angew. Chem. Int. Ed. Engl. {\bf 4}, 467 (1995) ; K. L. Taft and S. J. Lippard
, J. Am. Chem. Soc. {\bf 112}, 9629 (1990) ; R. W. Saalfrank, I. Bernt, 
E. Uller and F. Hampel, Angew. Chem. {\bf 109}, 2596 (1997) ; A. Caneschi,
A. Cornia, A. C. Fabretti and D. Gatteschi, Angew. Chem. Int. Ed. Engl. 
{\bf 38}, 1295 (1999).

\bibitem{fisher} J. C. Fisher and M. E. Fisher, Phys. Rev. {\bf 135}, A640
(1964).

\bibitem{haldane} F. D. M. Haldane, Phys. Lett. {\bf 93A}, 464 (1983) ; Phys.
Rev. Lett. {\bf 50}, 1153 (1983) ; I. Affleck, J. Phys. Condens. Matter. 
{\bf 1}, 3047 (1989).

\bibitem{gatteschi} D. Gatteschi and L. Pardi, Gazz. Chim. Ital. {\bf 123},
231 (1993).

\bibitem{waldmann} O. Waldmann, R. Koch, S. Schromm, J. Sch\"ulein, P. M\"uller
, I. Bernt, R. W. Saalfrank, F. Hampel and E. Balthes, Inorg. Chem. {\bf 40},
2986 (2001) ; O. Waldmann, Phys. Rev. B {\bf 61}, 6138 (2000).

\bibitem{silbey} Juan Bruno and R. J. Silbey, J. Phys. Chem. A {\bf 104}, 596
(2000).

\bibitem{cornia} A. Cornia, A. G. M. Jansen and M. Affronte, Phys. Rev. B
{\bf 60}, 12177 (1999).

\bibitem{julien} M. H. Julien, Z. H. Jang, A. Lascialfari, F. Borsa, M. 
Horvatic, A. Caneschi and D. Gatteschi, Phys. Rev. Lett. {\bf 83}, 227 (1999).

\bibitem{affronte} M. Affronte, J. C. Lasjaunias, A. Cornia and A. Caneschi,
Phys. Rev. B {\bf 60}, 1161 (1999).

\bibitem{loss1} A. Chilero, D. Loss, Phys. Rev. Lett. {\bf 80}, 169 (1998).

\bibitem{loss2} B. Normand, X. Wang, X. Zotos and D. Loss, Phys. Rev. B 
{\bf 63}, 184409 (2001).

\bibitem{raedt} S. Miyashita, K. Saito and H. De Raedt, Phys. Rev. Lett. 
{\bf 80}, 1525 (1998). 

\bibitem{gatteschi2} A. Caneschi , D. Gatteschi , C. Sangregorio , R. Sessoli,
L. Sorace , A. Cornia , M. A. Novak , C. Paulsen  and W. Wernsdorfer,
J. Magn. Magn. Mater. {\bf 200}, 182 (1999). 

\bibitem{vb} For a review see S. Ramasesha and Z.G. Soos, ``Valence Bond Theory
of Quantum Cell Models'', in {\it Valence Bond Theory and Chemical Structure}
Eds. D.J. Klein and D.L. Cooper, Elsevier (Amsterdam) in Press.

\bibitem{ramasesh} S. Ramasesha and Z. G. Soos, J. Chem. Phys. {\bf 98}, 4015
 (1993)

\bibitem{davidson} E.R. Davidson, J. Comput. Phys. {\bf 17}, 87 (1975).

\bibitem{zvezdin} A. K. Zvezdin, cond-mat/0004074 .

\bibitem{pearson} J. D. Cloizeaux and J. J. Pearson, Phys. Rev. {\bf 128}, 
2131 (1962).

\bibitem{floquet} J. H. Shirley, Phys. Rev. {\bf 138}, B979 (1965).

\end{thebibliography}
\end{document}